\documentclass[twocolumn,nofootinbib,prd,aps,prd,superscriptaddress]
{revtex4}

\usepackage{amsmath}
\usepackage{graphicx}
\usepackage{ifpdf}
\usepackage{comment}
\pdfoutput=1 

\def\lqcd{\Lambda_{\rm QCD}}

\newcommand{\nn}{\nonumber \\}
\newcommand\beq{\begin{equation}}
\newcommand\eeq{\end{equation}}

\def\be{\begin{equation}}
\def\ee{\end{equation}}
\def\ba{\begin{eqnarray}}
\def\ea{\end{eqnarray}}
\def\ge{\mathrel{\raise.3ex\hbox{$>$\kern-.75em\lower1ex\hbox{$\sim$}}}}
\def\la{\mathrel{\raise.3ex\hbox{$<$\kern-.75em\lower1ex\hbox{$\sim$}}}}
\def\eqn#1{Eq.~(\ref{#1})}
\def\simgt{\mathrel{\raise.3ex\hbox{$>$\kern-.75em\lower1ex\hbox{$\sim$}}}}
\def\simlt{\mathrel{\raise.3ex\hbox{$<$\kern-.75em\lower1ex\hbox{$\sim$}}}}

\newcommand{\nc}{\newcommand}

\nc{\gone}{\bar g_{\pi NN}^{(1)}}
\nc{\gzero}{\bar g_{\pi NN}^{(0)}}
\nc{\al}{\alpha}
\nc{\ga}{\gamma}
\nc{\de}{\delta}
\nc{\ep}{\epsilon}
\nc{\ze}{\zeta}
\nc{\et}{\eta}
\nc{\ka}{\kappa}
\nc{\rh}{\rho}
\nc{\si}{\sigma}
\nc{\ta}{\tau}
\nc{\up}{\upsilon}
\nc{\ph}{\phi}
\nc{\ch}{\chi}
\nc{\ps}{\psi}
\nc{\om}{\omega}
\nc{\Ga}{\Gamma}
\nc{\De}{\Delta}
\nc{\La}{\Lambda}
\nc{\Si}{\Sigma}
\nc{\Up}{\Upsilon}
\nc{\Ph}{\Phi}
\nc{\Ps}{\Psi}
\nc{\Om}{\Omega}
\nc{\ptl}{\partial}
\nc{\del}{\nabla}
\nc{\ov}{\overline}
\nc{\newcaption}[1]{\centerline{\parbox{15cm}{\caption{#1}}}}

\def\beq{\begin{equation}}
\def\eeq{\end{equation}}
\def\bmat{\begin{displaymath}}
\def\emat{\end{displaymath}}
\def\bea{\begin{eqnarray}}
\def\eea{\end{eqnarray}}
\def\ba{\begin{eqnarray}}
\def\ea{\end{eqnarray}}
\def\bery{\begin{array}}
\def\ery{\end{array}}
\def\bit{\begin{itemize}}
\def\eit{\end{itemize}}
\def\ben{\begin{enumerate}}
\def\een{\end{enumerate}}
\def\btab{\begin{tabular}}
\def\etab{\end{tabular}}
\def\btbl{\begin{table}}
\def\etbl{\end{table}}
\def\bfig{\begin{figure*}[htb]}
\def\efig{\end{figure*}}
\def\bpic{\begin{picture}}
\def\epic{\end{picture}}

\def\ga{\mathrel{\raise.3ex\hbox{$>$\kern-.75em\lower1ex\hbox{$\sim$}}}}
\def\la{\mathrel{\raise.3ex\hbox{$<$\kern-.75em\lower1ex\hbox{$\sim$}}}}
\def\gappeq{\mathrel{\rlap {\raise.5ex\hbox{$>$}}
{\lower.5ex\hbox{$\sim$}}}}
\def\lappeq{\mathrel{\rlap{\raise.5ex\hbox{$<$}}
{\lower.5ex\hbox{$\sim$}}}}

\def\gyr{{\rm \, G\kern-0.125em yr}}
\def\mev{{\rm \, Me\kern-0.125em V}}
\def\gev{{\rm \, Ge\kern-0.125em V}}
\def\tev{{\rm \, Te\kern-0.125em V}}

\def\msbar{\overline{\rm MS}}
\def\kt{k_\perp}
\def\OMIT#1{{}}

\def\fig#1{Fig.\ \ref{#1}}

\def\vereq#1#2{\lower3pt\vbox{\baselineskip1pt\lineskip1pt
     \ialign{\nonumber \\$#1\hfill##\hfil\nonumber \\$\crcr#2\crcr\sim\crcr}}}

\makeatletter
\def\fmslash{\@ifnextchar[{\fmsl@sh}{\fmsl@sh[0mu]}}
\def\fmsl@sh[#1]#2{%
   \mathchoice
     {\@fmsl@sh\displaystyle{#1}{#2}}%
     {\@fmsl@sh\textstyle{#1}{#2}}%
     {\@fmsl@sh\scriptstyle{#1}{#2}}%
     {\@fmsl@sh\scriptscriptstyle{#1}{#2}}}
\def\@fmsl@sh#1#2#3{\m@th\ooalign{$\hfil#1\mkern#2/\hfil$\crcr$#1#3$}}

\makeatother
\def\JADE{\rm{JADE}}
\def\SW{\rm{SW}}
\def\soft{s}
\def\ktp{k_3^+}
\def\ktm{k_3^-}

\def\ptm{p_3^-}
\def\lambdaf{\Lambda_f}

\begin{document}
\ifpdf
\DeclareGraphicsExtensions{.pdf, .jpg}
\else
\DeclareGraphicsExtensions{.eps, .jpg}
\fi


\title{Phase Space and Jet Definitions in SCET}
\author{William Man-Yin Cheung}
\email{mycheung@physics.utoronto.ca}
\affiliation{Department of Physics, University of Toronto, 
     60 St.\ George Street, Toronto, Ontario, Canada M5S 1A7}
\author{Michael Luke}
\email{luke@physics.utoronto.ca}
\affiliation{Department of Physics, University of Toronto, 
     60 St.\ George Street, Toronto, Ontario, Canada M5S 1A7}
\author{Saba Zuberi}
\email{szuberi@physics.utoronto.ca}
\affiliation{Department of Physics, University of Toronto, 
     60 St.\ George Street, Toronto, Ontario, Canada M5S 1A7}\date{\today}

\begin{abstract}
We discuss consistent power counting for integrating soft and collinear degrees
of freedom over arbitrary regions of phase space in the soft-collinear effective
theory (SCET), and illustrate our results at one loop with several jet
algorithms:  JADE, Sterman-Weinberg and $\kt$.  Consistently applying SCET
power-counting in phase space, along with non-trivial zero-bin subtractions,
prevents double-counting of final states.  The resulting phase-space integrals
over soft and collinear regions are individually ultraviolet divergent, but the
phase-space ultraviolet divergences cancel in the sum.   Whether the soft and
collinear contributions are individually infrared safe depends on the jet
definition.  We show that while this is true at one loop for JADE and
Sterman-Weinberg, the $\kt$ algorithm does not factorize into individually
infrared safe soft and collinear pieces in dimensional regularization.  We point
out that this statement depends on the ultraviolet regulator, and that in a
cutoff scheme the soft functions are infrared safe.

\end{abstract}

\maketitle

\section{Introduction}

The study of jets provides an important tool to investigate strong interactions
and tests QCD over a wide range of scales, from partonic hard scattering to the
evolution of hadronic final states that make up the jets. Hadronic jets also
play an integral role in searches for physics beyond the Standard Model.
Soft-collinear effective theory (SCET) \cite{Bauer:2000ew, Bauer:2000yr,
Bauer:2001ct, BPS,Bauer:2002nz} provides a useful framework to study jets, reproducing
results from QCD obtained from traditional factorization techniques
(see, for example, \cite{Collins:1989gx,Sterman:1995fz})
while systematically including power corrections and organizing perturbative
resummation. 

The effective theory separates the scales of the underlying hard interaction
from the scales associated with the collinear particles in the jets and the
long-distance soft physics. Unlike QCD, particles in SCET whose momenta have
parametrically different scaling are described by separate fields - in this
case, either \mbox{(ultra-)soft} or collinear\footnote{In situations with
multiple collinear directions, there are collinear modes for each direction.}.
Their light-cone components, $p=(n\cdot p,\bar{n} \cdot p,
p_\perp)=(p^+,p^-,p^\perp)$ scale as :
\begin{equation}
 p_s\sim Q(\lambda^2, \lambda^2, \lambda^2), \ p_c\sim Q(1, \lambda^2, \lambda)
\end{equation}
where $n$ and $\bar{n}$ are light-cone vectors in the $\pm \vec{n}$ direction
and $\lambda$ is a small dimensionless parameter which is determined by the
dynamics. At leading order in $\lambda$ the soft and collinear modes decouple in
the SCET Lagrangian. These properties of the effective theory have been utilized
to prove factorization, resum large logarithms and parameterize nonperturbative
corrections for event shapes in the two jet limit \cite{short2jet, long2jet,
Bauer:2008dt, Hornig:2009vb} and for massive top quark jets
\cite{Fleming:2007qr}, for example. The factorization of generic fully
differential jet cross sections has also been shown independent of jet
observables for $\mathit{e}^+ \mathit{e}^-$ and $\mathit{p} \mathit{p}$
collisions \cite{jetfactor}.  For an $n$-jet cross section with a given jet
definition to fully factorize, however, the phase space constraints should also
factorize appropriately in the effective field theory (EFT). Such factorization
of phase space constraints has not yet been shown in any scheme other than the
hemisphere scheme \cite{jetfactor} (in which all events are dijet).

In this paper we study the two-jet cross section for $\mathit{e}^+ \mathit{e}^-$
collisions in SCET, using three jet algorithms:  a cone algorithm,
Sterman-Weinberg (SW) \cite{SW}, which defines a jet based on an angularity cut
and was considered in the context of SCET in \cite{short2jet, long2jet, Trott}, as well as two
clustering algorithms, JADE \cite{JADE} and $\kt$ \cite{Catani:1991hj}, which
iteratively combine partons into jets based on kinematic conditions.   This is a
first step towards the broader goal of determining the appropriate factorization
theorem and resumming logarithms using SCET.   While we do not consider here the
more general problem of factorization theorems for jets, we point out some
implications of our results for factorization theorems, in particular showing that the form
of the factorization in SCET depends on the ultraviolet regulator.
The main point of this paper is instead to demonstrate the relationship
between the cutoffs in the effective field theory and phase space limits, and to
consider their implications for dijet rates in SCET.  Since SCET has no hard
cutoff separating soft from collinear regions of phase space, some care is
required to perform phase space integrals consistently.  
The NLO dijet rate in SCET also demonstrates the interplay of
divergences between the soft and collinear sectors, and provides nontrivial
examples of the zero-bin subtraction \cite{zerobin}.

\section{Phase Space in QCD and SCET}
\label{ps}

\begin{figure}[thb]
\centering 
\includegraphics[width=7.5cm]{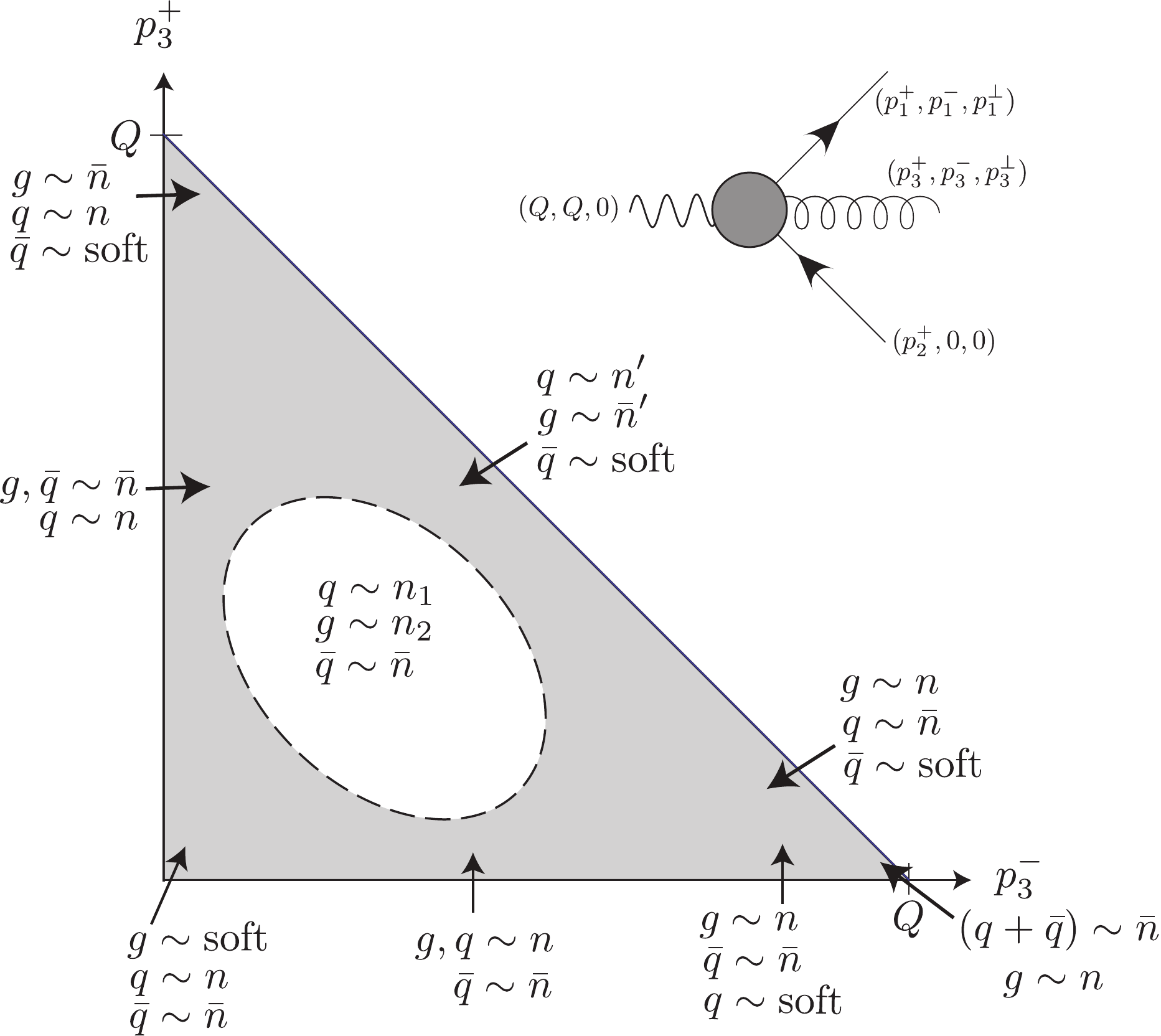}
\caption{Three-body phase space in $p_3^+$, $p_3^-$ variables.  The shaded area
indicates regions which may be described with two collinear directions in SCET;
the white region in the centre requires three directions.}\label{kpkmphasespace}
\end{figure}

At each order  in perturbation theory, a jet algorithm corresponds to a scheme
to partition the available phase space into regions with different numbers of
jets.  At $O(\alpha_s)$, the phase space for $e^+ e^-\to\mbox{hadrons}$ or
hadronic $Z$ decay was discussed in SCET in \cite{zerobin} using the variables
$x_i={2p_i\cdot q\over q^2}$, where $q=p_1+p_2+p_3$ is the total momentum of the
process and $p_{1,2,3}$ are the momenta of the quark, antiquark and gluon,
respectively.  In our discussion we will find it more convenient to choose the
independent variables to be the light-cone components of the gluon momentum,
$p_3^+\equiv n\cdot p_3$ and $p_3^-\equiv\bar n\cdot p_3$, and fix the
coordinates by choosing the antiquark to be moving purely in the $\bar n$
direction (i.e. $p_2^-=p_2^\perp=0$).  The resulting phase space is illustrated
schematically in \fig{kpkmphasespace}.
Note that because our choice of coordinates is not symmetric in the $n$ and
$\bar n$ directions, the phase space is not symmetric under exchange of the
$p_3^+$ and $p_3^-$ axes.  (For example, in the upper left the antiquark is constrained
to be soft, while in the lower right the quark and antiquark recoil against the gluon,
and so either the quark or the antiquark may be soft, or both may be $\bar n$-collinear.)

In the shaded regions, two of the partons recoil approximately back-to-back and
the third is either soft or recoils roughly parallel with one of the other two,
while in the central unshaded region all three partons recoil in different
directions.  Thus, the shaded region roughly corresponds to two-jet events,
while the central region corresponds to three-jet events.  The precise details
of this correspondence are determined by the particular jet algorithm being
used.

Within the effective field theory there are natural degrees of freedom
associated with each region of the two jet phase space, as indicated in
\fig{kpkmphasespace}.   The complete dijet rate, however, requires integrating
over all these regions, and since SCET has no hard cutoff separating soft and
collinear degrees of freedom, it would seem that each mode should be integrated
over the full QCD phase space (this is the approach followed in \cite{zerobin}).
 However, this is inconsistent with the effective theory, since, for example,
integrating a soft gluon in the collinear region would require it to have
momentum well above the cutoff for soft modes in SCET.

\begin{figure}[htb]
\centering 
\includegraphics[width=8cm]{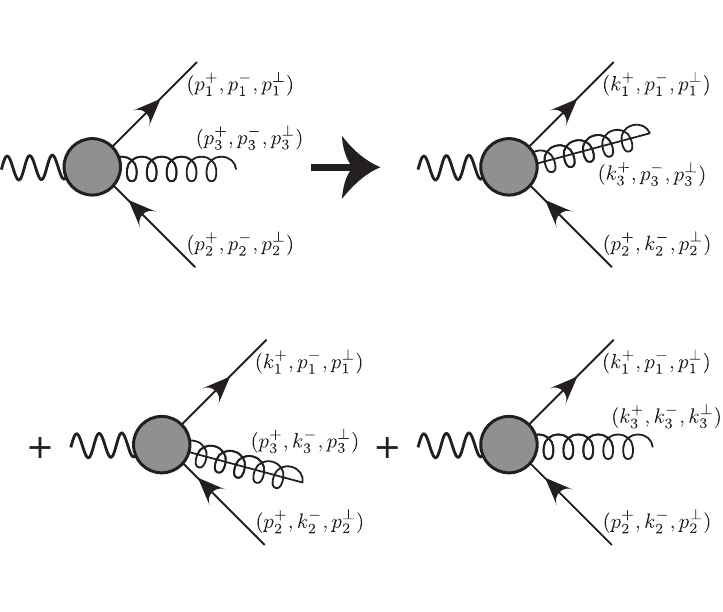}
\caption{Kinematics in SCET.  In the first SCET diagram the gluon is
$n$-collinear, in the second it is $\bar n$-collinear, and in the third it is
soft.  Additional diagrams with soft quarks arise at higher order in
$\lambda$.}\label{kinematicsfig}
\end{figure}

Instead, a phase space integral which extends above the cutoff for the relevant
mode should be replaced by an ultraviolet divergence, which would then be
regulated and renormalized in the usual way.  This occurs naturally in SCET
because of the multipole expansion for momenta at the vertices.  The kinematics
for soft and collinear gluon emission is shown in \fig{kinematicsfig}, where
$p^{\pm}$ scale as $Q$, $p^\perp$ scale as $\lambda Q$ and the $k$'s scale as
$\lambda^2 Q$.   Because of the multipole expansion, a given component of
momentum is not conserved at vertices involving particles whose typical momenta
scale differently with $\lambda$.  As a consequence, the phase space for each
mode in SCET differs from that in full QCD, and it is misleading to use the
kinematics in \fig{kpkmphasespace} in the effective theory.   For example, in
the soft emission graph in \fig{kinematicsfig}, conservation of momentum
requires $p_1^-=Q$, $p_2^+=Q$, while the $k$'s are unconstrained.  It is
integrals over these unconstrained momenta which will give rise to ultraviolet
divergent phase space integrals in the EFT.   This is the approach followed in
\cite{ Hornig:2009vb}, where ultraviolet divergent phase space integrals are
obtained for the soft and jet functions at NLO for angularity distributions in
SCET. This is also what happens in SCET in loop graphs, where both soft and
collinear degrees of freedom propagate, integrated over the appropriate
kinematic variables.  Since phase space integrals are just loop graphs with
internal propagators placed on shell, the same rules apply.

\begin{figure*}[htb]
\centering 
\includegraphics[width=\textwidth]{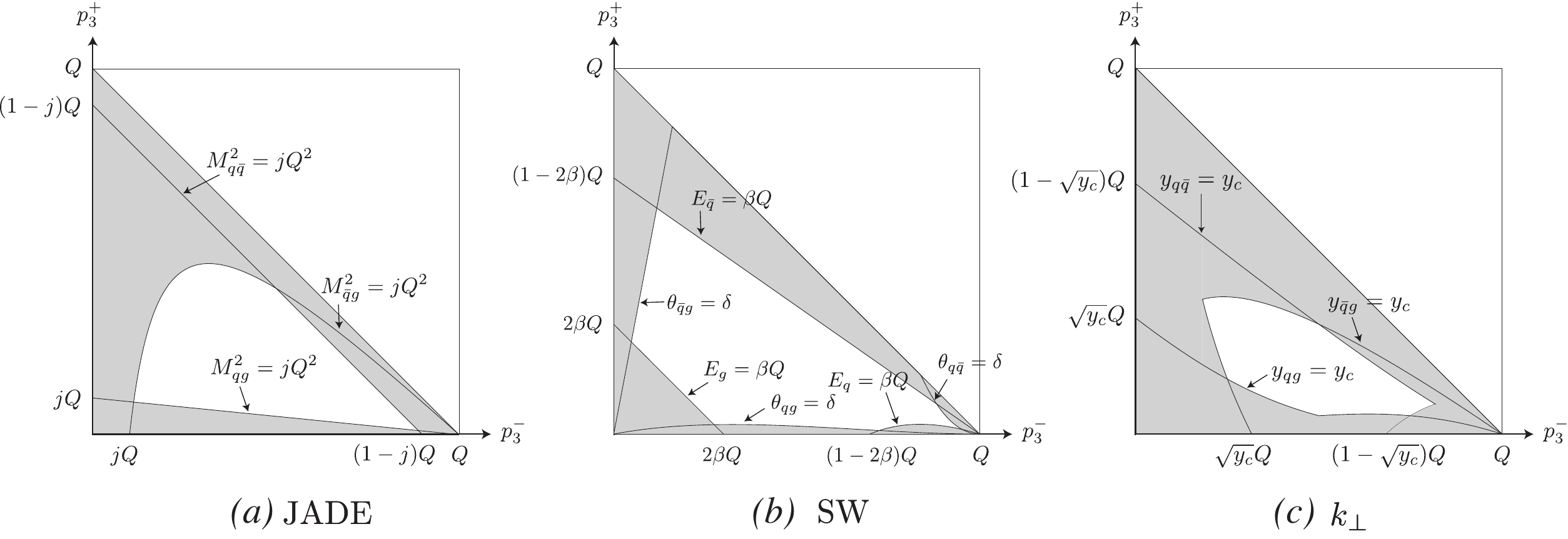}
\caption{Three-body phase space for different jet definitions in QCD.  The
shaded region corresponds to the two jet region; the unshaded region in the
centre is the three-jet region.}\label{jetdefinitionsPS}
\end{figure*}

It is straightforward to illustrate this for various jet definitions.  In the SW
definition, a two-jet event is defined as one in which all but a fraction
$\beta$ of the total energy of the event is deposited in two back-to-back cones
with half angle $\delta$ \cite{SW}.   The JADE algorithm requires that the
invariant mass $M_{ik}^2$ of every pair of final-state partons $i$ and $k$ be
calculated. If any are less than a fraction, $j$, of the total center of mass
energy squared, $Q^2$, then the momenta of the pair with the smallest invariant
mass are combined into a single jet according to a recombination scheme which is
part of the jet definition, the details of which are not relevant at
$O(\alpha_s)$. This process is repeated until no pair has an invariant mass less
than $j Q^2$. The $\kt$ algorithm is a modified version of the JADE algorithm
which clusters partons based on their relative transverse momentum rather than
their invariant mass.  The corresponding kinematic variable is
\bea \label{kt}
y_{ij} =  \frac{2}{Q^2} (1-\cos \theta_{ij}) \, \mathrm{min}
\left(E_i^2,E_j^2\right).
\eea
For massless particles this is equal to 
\bea
y_{ij} =  \frac{M_{ij}^2}{Q^2}  \,  \mathrm{min} \left(\frac{E_i}{E_j},
\frac{E_j}{E_i} \right).
\eea
The final states with the smallest $y_{ij}$, given that it is less than a
resolution parameter $y_c$, are combined according to a combination
prescription. This process is repeated until all pairs have $y_{ij}>y_c$.  In
\fig{jetdefinitionsPS} we illustrate the two-jet regions in QCD as defined by
the JADE, SW and $\kt$ algorithms.   The SCET regime for the two-jet cross
section corresponds to choosing the parameters $\delta$, $\beta$, $j$ or $y_c$
to be much less than one in the respective jet definition.

For the two jet JADE cross section, for example, integrating $k_3^+$ 
in the soft sector all the way up to $Q$, as in
\fig{jetdefinitionsPS}(a), corresponds to integrating the gluon momentum far
above the cutoff.  In the EFT, the upper limit of integration should therefore
be replaced by an ultraviolet cutoff.  Indeed, while the regions of integration
for the various jet definitions are quite complicated, as far as the soft gluon
is concerned they should have no structure above the soft scale.  A similar
situation holds for collinear gluons, where the effective cutoffs in the
perpendicular and anti-collinear directions are parametrically smaller than $Q$.

\begin{figure*}[p]
\centering 
\includegraphics[width=12cm]{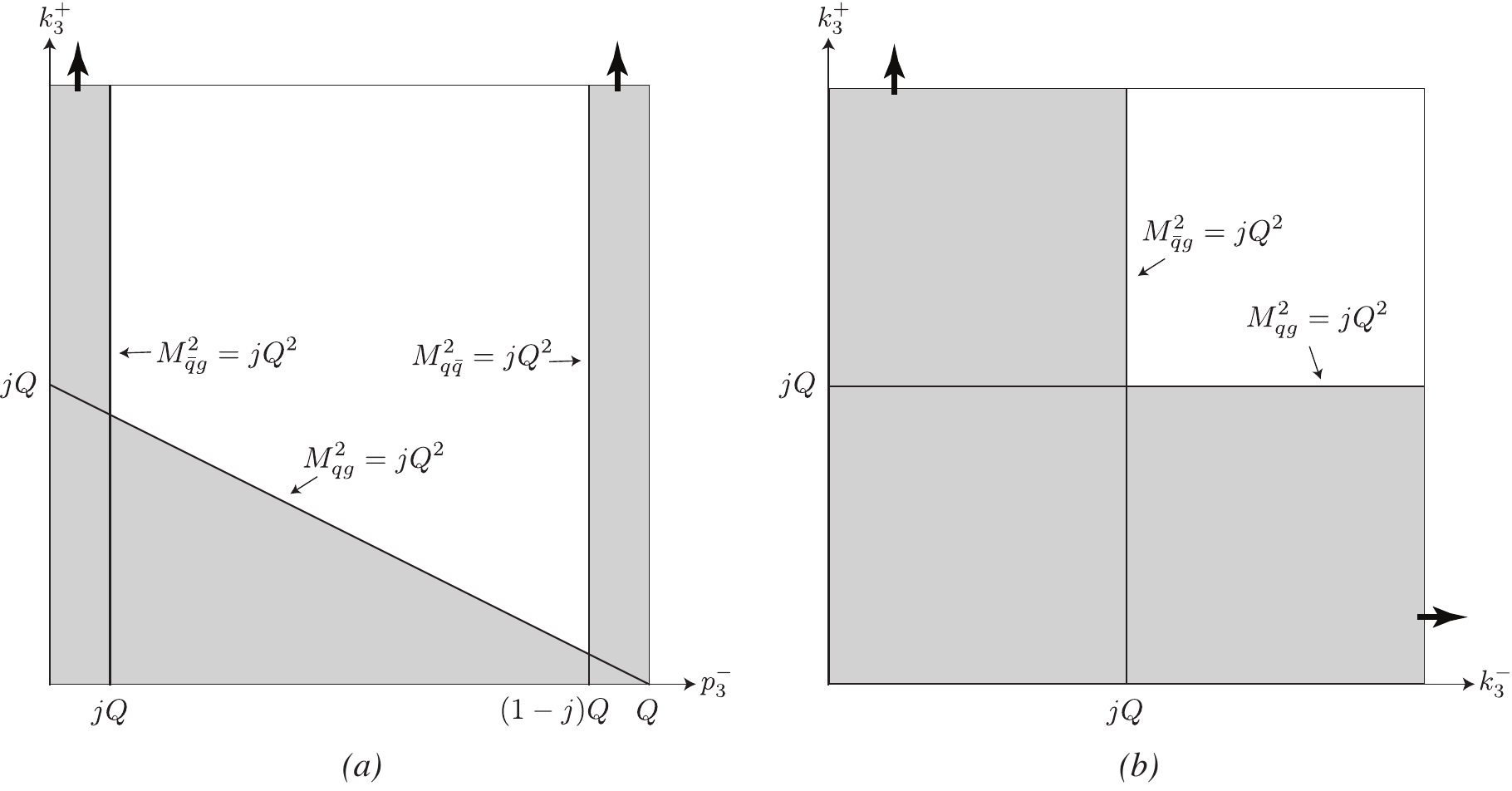}
\caption{Phase space corresponding to two-jet events using the JADE algorithm in
(a) the $n$-collinear gluon sector, and (b) the soft gluon and zero-bin sectors.
 The thick arrows indicate integrations to infinity.   }\label{jadeplotsfig}
\end{figure*}

At $O(\alpha_s)$, the JADE algorithm corresponds to a cut on the invariant
masses $M_{ij}$ of each pair of partons:  if $M_{ij}^2< jQ^2$, the partons are
considered to lie in the same jet, and the event is a two-jet event.  The
constraints in full QCD shown in \fig{jetdefinitionsPS}(a) are
\begin{eqnarray}
&&\frac{M_{qg}^2}{Q^2}={p_3^+\over Q-p_3^-}<j, \quad \frac{M_{\bar
qg}^2}{Q^2}=\frac{p_3^-}{Q}-\frac{p_3^+ p_3^-}{Q(Q-p_3^-)}<j, \nn
&&\frac{M_{q\bar q}^2}{Q^2}=\frac{Q-p_3^--p_3^+}{Q}<j.
\end{eqnarray}
Expanding these constraints in the $n$-collinear sector, we find
\begin{eqnarray}
\label{jadeNconst}
&&\frac{M_{qg}^2}{Q^2} =\frac{\ktp}{Q-\ptm} < j , \quad\frac{M_{\bar q
g}^2}{Q^2} = \frac{\ptm}{ Q} < j , \nn
&&\frac{M_{q\bar q}^2}{Q^2} = \frac{Q-\ptm}{Q}<j
\end{eqnarray}
while in the soft sector we obtain
\begin{equation}\label{jadeSconst}
\frac{M_{qg}^2}{Q^2} = \frac{\ktp}{Q} < j , \quad \frac{M_{\bar q g}^2}{Q^2} =
\frac{\ktm}{ Q} < j
\end{equation}
(while the constraint $M_{q\bar q}^2< j Q^2$ is never satisfied).
Finally, in order to avoid double-counting of the soft sector, the zero-bin of
the collinear region must be subtracted \cite{zerobin}.  Taking the soft limit
of the $n$-collinear region in \eqn{jadeNconst} gives the same region as the
soft sector, \eqn{jadeSconst}.  The corresponding regions of phase-space are
shown in \fig{jadeplotsfig}(a, b).

\begin{figure*}[p]
\centering 
\includegraphics[width=\textwidth]{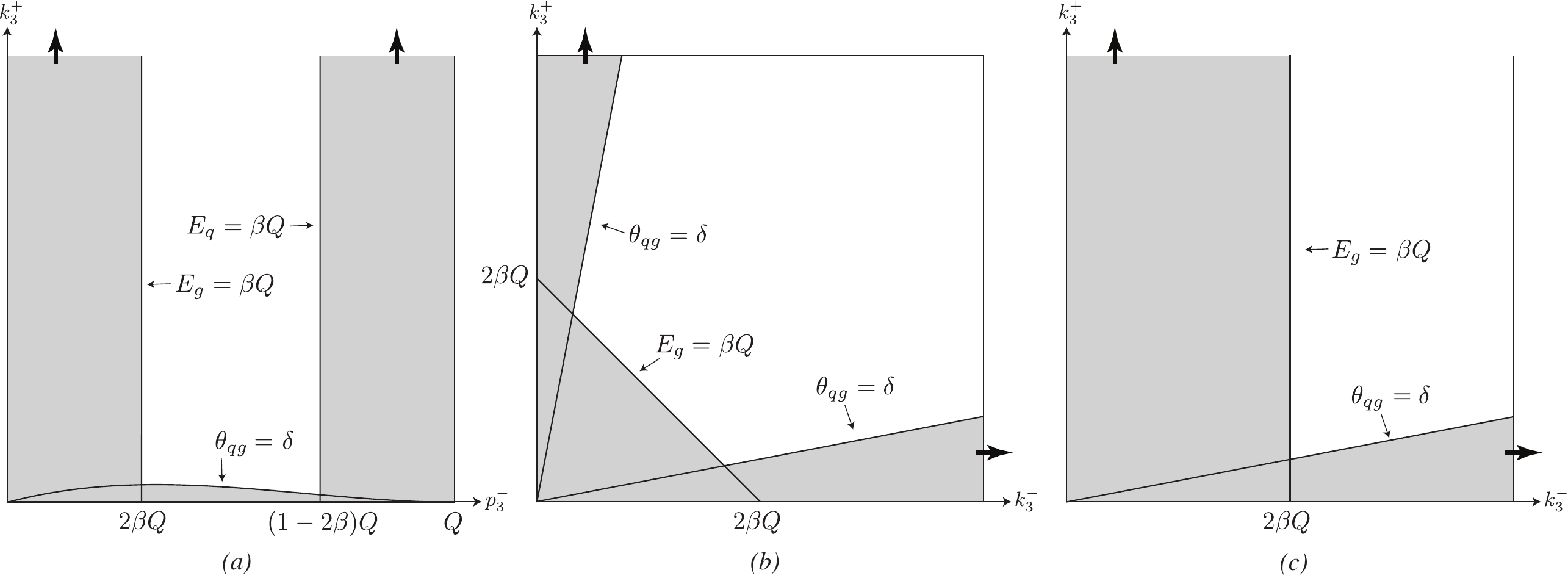}
\caption{Phase space corresponding to two-jet events using the SW algorithm in
(a) the $n$-collinear gluon sector, (b) the soft gluon sector, and (c) the
zero-bin sector.  As before, the thick arrows indicate integrations to infinity.
  }\label{swplotsfig}
\end{figure*}

\begin{figure*}[p]
\centering 
\includegraphics[width=\textwidth]{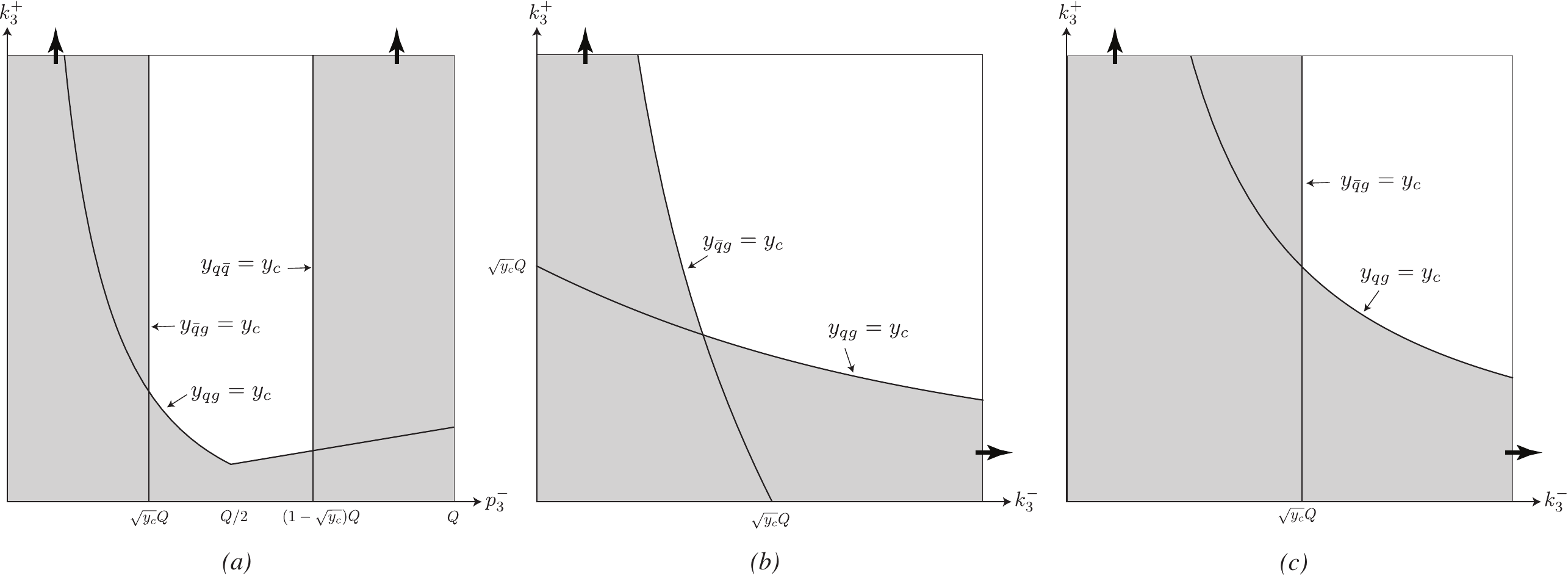}
\caption{As \fig{swplotsfig}, but using the $\kt$ algorithm.}\label{ktplotsfig}
\end{figure*}

\begin{table*}[t]
\begin{tabular}{cccc}
Jet Definition  
   & ~$n$-collinear regions~ &  ~soft regions~ & ~zero-bin regions~     \\
\hline\hline
JADE &~$\ktp <  j(Q-\ptm)$~&~$ \ktp<j Q$~&~$\ktp<j Q$~  \\
&~$\ptm<jQ$ ~&~ $ \ktm<j Q$~&~$ \ktm<j Q$~\\
&~$\ptm>Q(1-j)$~&~  ~&  \\   \hline
SW &~$\ktp <  \ptm  \frac{(Q -\ptm)^2}{Q^2} \delta^2$~&~$ \frac{\ktp}{\ktp +
\ktm}< \delta^2$~&~$\ktp < \delta^2 \ptm$~  \\
&~$\ptm<2\beta Q$ ~&~ $ \frac{\ktm}{\ktp + \ktm}<\delta^2$~&~$ \ptm < 2\beta Q$~
 \\
&~$\ptm>(1-2\beta)Q$~&~ $\ktp + \ktm<  2\beta  Q$ ~&  \\   \hline
$\kt$& ~$\vphantom{\Bigg(} \mathrm{min} \left(\frac{\ktp}{\ptm}, \frac{\ktp
\ptm}{(Q-\ptm)^2}\right)<y_c $~&~~$\left(\ktp +\ktm\right) \ktp<y_c Q^2$~~&~$
\ktp \ptm<y_c Q^2$~ \\
&~$(\ptm)^2<y_c Q^2$~&~$\left(\ktp +\ktm\right) \ktm<y_c Q^2$~&~$(\ptm)^2<y_c
Q^2$~\\
&~$(Q-\ptm)^2<y_c Q^2$~&&\\ \hline
\end{tabular}\vspace*{4pt}
\caption{Two-jet regions of three-body phase space for JADE, Sterman-Weinberg
(SW) and $\kt$ jet algorithms.}
\label{jettable}
\end{table*}

We note that, as required, the phase space contains no explicit reference to any
scales above the cutoff of the theory and has no structure above this scale. 

Similar constraints in the soft, collinear and zero-bin sectors are easily
obtained for the SW and $\kt$ definitions, and are summarized in Table
\ref{jettable}.  Note that in both SW and $\kt$, the zero-bin region is not the
same as the soft region, since taking the soft limit of the $n$-collinear phase
space is not the same as taking the soft limit of the full QCD phase space.  The
corresponding regions are illustrated in Figs. \ref{swplotsfig} and
\ref{ktplotsfig}.

Note that we have not had to specify the SCET expansion parameter $\lambda$
in order to expand the phase space in the soft and collinear sectors; we have only
assumed that $\lambda\ll 1$ so that the multipole expansion is valid.   Similarly, we
have not assumed any relative scaling between $\beta$ and $\delta$ in the SW jet
definition.

\section{Dijet Rates to $O(\alpha_s)$}

In this section we calculate the NLO dijet rate (denoted $f_2$) in the JADE, SW
and $\kt$ schemes in SCET,  which is straightforward to do given the phase space
regions of the previous section.  We show that in each case SCET reproduces full
QCD, as it must. We examine the scales that appear in the soft and collinear
cross sections, where the power counting parameter $\lambda$ is determined by
the dynamics in each algorithm. It is instructive to note the cancellation of
ultraviolet divergences between the soft and collinear real emission
contributions. We also  consider the infrared safety of the soft and collinear
rates separately. 

At $O(\alpha_s)$ the only contribution to the dijet rate comes from the two-jet
SCET operator $O_2 =
\bar{\xi}_nW_n\gamma^{\mu}W^\dagger_{\bar{n}}\xi_{\bar{n}}$.  The matching
calculation from the full QCD current $\bar{\psi}\gamma^{\mu}\psi$ onto $O_2$
has been performed many times in the 
literature \cite{dis,Trott, BauerSchwartz}, with the Wilson coefficient 
\begin{equation}\label{c2}
C_2 = 1+\frac{\alpha_s C_F}{2 \pi} \left(-\frac{1}{2} \ln^2
\frac{\mu^2}{-Q^2}-\frac{3}{2} \ln \frac{\mu^2}{-Q^2} -4 + \frac{ \pi^2}{12}
\right)
\end{equation}
and the $\msbar$ counterterm
\begin{equation}\label{z2}
Z_2 = 1+\frac{\alpha_s C_F}{2 \pi} \left(\frac{1}{\epsilon^2}+\frac{3}{2
\epsilon}+\frac{1}{\epsilon} \ln \frac{\mu^2}{-Q^2} \right)
\end{equation}
where we are working in $d=4-2 \epsilon$ dimensions.
The SCET differential cross section for soft gluon emission is given by
\begin{equation}\label{softDiffDR}
\frac{1}{\sigma_0}d \sigma^\soft = \frac{\alpha_s C_F}{2 \pi}  \frac{\mu^{2
\epsilon} e^{\epsilon \gamma_E} }{\Gamma(1-\epsilon)} dk_3^+ dk_3^- \frac{2 \,
\theta (k_3^+ k_3^-)}{(k_3^+)^{1+\epsilon}(k_3^-)^{1+\epsilon} }
\end{equation}
while for $n$-collinear gluon emission it is
\begin{eqnarray} \label{nDiffDR}
\frac{1}{\sigma_0}d \sigma^{n} &=& \frac{\alpha_s C_F}{2 \pi}  \frac{\mu^{2
\epsilon} e^{\epsilon \gamma_E} }{\Gamma(1-\epsilon)} d\ktp d\ptm \frac{(\ptm
\ktp)^{- \epsilon}}{Q \ktp} \nn
&&\times \left( \frac{\ptm}{Q} (1-\epsilon)+2 \frac{Q-\ptm}{\ptm} \right)
\eea
where $\sigma_0 = (4\pi\alpha^2/Q^2)\sum_f e_f^2$ is the leading order Born
cross section with a sum over the (anti-)quark charges $e_f$.  The dependence on
$\vec{k}_3^\perp$ and $\vec p_3^\perp$ has been eliminated via the gluon
on-shell condition, and the integral over the $2-2\epsilon$ perpendicular
components of the gluon momentum has been performed in each case.

Finally, the differential rate in the gluon zero-bin region, $d\sigma^{n0}$, is
obtained by taking the soft limit of \eqn{nDiffDR}, which is the same as the
soft rate,
\begin{equation}\label{zerobinDrate}
d\sigma^{n0}=d\sigma^\soft.
\end{equation}
(There are also zero-bin regions corresponding to the quark and antiquarks becoming
soft, but they are higher order in $\lambda$ and we will not consider them
here.)
For the $n$-collinear region there are two zero-bins: $\ptm \to 0$ and $p_1^-
\to 0$, but the contribution to the cross section from the latter is of higher
order in $\lambda$ and so we will not consider them here.

\subsection{JADE}\label{JADEsection}

Integrating the soft rate over the soft dijet region (\ref{jadeSconst}) in the
JADE definition gives
\begin{eqnarray}\label{JADEsoftPDR}
&&\frac{1}{\sigma_0}\sigma_{\JADE}^{\soft} \nn
&=& \frac{\alpha_s C_F}{2 \pi} \left( - \frac{2}{\epsilon^2} -\frac{2}{\epsilon}
\ln \frac{\mu^2}{j ^2 Q^2}-\ln^2\frac{\mu^2}{j ^2 Q^2}+ \frac{\pi^2}{6}   
\right)\nn
\end{eqnarray}
where we have taken $j\ll 1$ and kept only the leading terms in $j$.   
Integrating the $n$-collinear rate over the region (\ref{jadeNconst}), we find 
\begin{eqnarray}\label{jadeN}
\frac{1}{\sigma_0}\tilde\sigma_{\JADE}^{n}
&=& \frac{\alpha_s C_F}{2 \pi} \left(  \frac{3}{2 \epsilon} +\frac{2}{\epsilon}
\ln j+ \frac{3}{2} \ln \frac{\mu^2}{j Q^2}\right.\nn
&&\left.+2 \ln\frac{\mu^2}{Q^2} \ln j- 3\ln^2 j - \frac{\pi^2}{3}+\frac{7}{2}
\right)
\end{eqnarray}
where the tilde denotes that the zero-bin has not been subtracted.  The rate in
the zero-bin region is identical to that in the soft region, and so the zero-bin
subtracted result for the emission of an $n$-collinear gluon is
\begin{eqnarray}
\frac{1}{\sigma_0}\sigma_{\JADE}^n &=&
\frac{1}{\sigma_0}(\tilde\sigma_{\JADE}^{n}-\sigma_{\JADE}^{n0}) =
\frac{1}{\sigma_0}(\tilde\sigma_{\JADE}^{n}-\sigma_{\JADE}^\soft)\nn 
&=& \frac{\alpha_s C_F}{2 \pi} \left(  \frac{2}{\epsilon^2}+\frac{3}{2 \epsilon}
+\frac{2}{\epsilon} \ln \frac{\mu^2}{j Q^2}+ \frac{3}{2} \ln\frac{\mu^2}{j
Q^2}\right.\nn
&&\left.+\ln^2 \frac{\mu^2}{j Q^2}  - \frac{\pi^2}{2}+\frac{7}{2} \right).
\end{eqnarray}
The emission of a collinear gluon in the $\bar{n}$ direction, i.e. from the
anti-quark, can be calculated in a similar way, and it gives the same
contribution.

In pure dimensional regularization, all the virtual vertex corrections and the
wavefunction renormalizations involve scaleless integrals and thus vanish. Hence
we only need to add up the real emission contributions:
\begin{eqnarray}\label{JADEsum}
&&{1\over\sigma_0}\sigma^R_{\JADE}\nn
&=&\frac{1}{\sigma_0}\left((\tilde\sigma_{\JADE}^{n}-\sigma_{\JADE}^{n0})+
(\tilde\sigma_{\JADE}^{\bar{n}}-\sigma_{\JADE}^{\bar{n}0})+\sigma_{\JADE}^{\soft
}\right)  \nn
 &=& \frac{1}{\sigma_0}(\tilde\sigma_{\JADE}^{n}+ \tilde\sigma_{\JADE}^{\bar{n}}
- \sigma_{\JADE}^{\soft})\nn
 &=& \frac{\alpha_s C_F}{2 \pi} \left( \frac{2}{\epsilon^2}+ \frac{3}{\epsilon}
+\frac{2}{\epsilon} \ln \frac{\mu^2}{Q^2} -2 \ln^2 j +\ln^2 
\frac{\mu^2}{Q^2}\right.\nn
 &&\left. + 3 \ln  \frac{\mu^2}{j Q^2} - \frac{5 \pi^2}{6}+7 \right).
\end{eqnarray}
Note that the soft contribution enters into the final expression with a minus
sign. This is a consequence of zero-bin subtraction and the fact that zero-bins
are identical to the soft contribution. Similar observations have been pointed
out in \cite{IdilbiMehen1, IdilbiMehen2, DeltaRegulator}. The divergent terms in
\eqn{JADEsum} are cancelled by the counter term $|Z_2|^2$, and including the
Wilson coefficient,  $|C_2|^2$, gives the two-jet fraction
\begin{eqnarray}\label{jadeQCD}
f_2^{\JADE}&=&
\frac{|C_2|^2}{|Z_2|^2}\left(1+{1\over\sigma_0}\left(\sigma_{\JADE}^{n}+
\sigma_{\JADE}^{\bar{n}} + \sigma_{\JADE}^{\soft}\right)\right)\nn
&=&1 + \frac{\alpha_s C_F}{2 \pi} \left(-2 \ln^2 j-3 \ln j+ \frac{\pi^2}{3}-1
\right).\nn
\end{eqnarray}
This result agrees with the full QCD calculation given in \cite{TwoJetJADE,
BrownStirling2}.

It is instructive to comment on a few details of the SCET result.
First of all, since dimensional regularization regulates both the infrared and
ultraviolet divergences, the cancellation of ultraviolet divergences between the
soft and collinear emissions is not explicit. To show how this works, we can
repeat the calculation with the quark and anti-quark offshell, $p_1^2,
p_2^2\sim\lambda^2\ne 0$, so that all $1/\epsilon$ divergences in the
calculation are ultraviolet.
The calculation is given in Appendix \ref{appendixa}. The resulting rate for
soft gluon emission over the JADE phase space is
\begin{eqnarray}\label{JADEsOS}
\frac{1}{\sigma_0}\sigma_{\JADE}^{\soft} &=& \frac{\alpha_s C_F}{2 \pi} \left(-
\frac{2}{\epsilon} \left(\ln \frac {p_1^2}{jQ^2}+ \ln \frac{p_2^2}{jQ^2} \right)
\right.\nn
&& + \left(\ln \frac {p_1^2}{Q^2} + \ln \frac{p_2^2}{Q^2} \right)^2\nn
&& -\left. 2 \left(\ln \frac {p_1^2}{Q^2} + \ln \frac{p_2^2}{Q^2} \right) \ln
\frac{\mu^2}{Q^2}\right)+\cdots\nn
\end{eqnarray}
where the ellipses denote finite constant terms which are not relevant for the
discussion.  The unsubtracted $n$-collinear cross section is
\begin{eqnarray}\label{jadeNtildeos}
\frac{1}{\sigma_0}\tilde\sigma_{\JADE}^{n}&=& \frac{\alpha_s C_F}{2 \pi}
\left(-\frac{2}{\epsilon^2}+  \frac{2}{\epsilon} \left(\ln \frac{p_1^2}{jQ^2}-
\ln \frac{\mu^2}{j^2Q^2} \right) \right.\nn
&&\left.-\ln^2 \frac{p_1^2}{Q^2} + 2 \ln\frac{\mu^2}{Q^2}\ln \frac{p_1^2}{Q^2} +
\frac{3}{2} \ln \frac{p_1^2}{Q^2}\right)\nn
 &&+ \dots
\end{eqnarray}
while the zero-bin region gives
\bea
\frac{1}{\sigma_0}\sigma_{\JADE}^{n0} = \frac{\alpha_s C_F}{2 \pi}
\left(-\frac{2}{\epsilon^2}-  \frac{2}{\epsilon} \ln \frac {\mu^2}{j^2 Q^2}
\right) +\dots \ .
\eea
Thus, the zero-bin subtracted $n$-collinear cross section is
\begin{eqnarray} \label{JADEnOS}
\frac{1}{\sigma_0}\sigma_{\JADE}^{n} &=& \frac{\alpha_s C_F}{2 \pi}
\left(\frac{2}{\epsilon} \ln \frac{p_1^2}{jQ^2} -\ln^2
\frac{p_1^2}{Q^2}\right.\nn
&&+\left.   2 \ln\frac{\mu^2}{Q^2}\ln \frac{p_1^2}{Q^2}+ \frac{3}{2} \ln
\frac{p_1^2}{Q^2}\right) + \dots\ .\nn
\end{eqnarray}
The result for  $\bar{n}$-collinear gluon emission will be the same as that for
$n$-collinear gluon emission with the replacement $p_1^2 \to p_2^2$.
Note that the $1/\epsilon^2$ divergence from collinear emission is removed by
the zero-bin. Combining the real emission contributions to the JADE cross
section, \eqn{JADErealOS}, we see that while the phase-space integrals for soft
and collinear gluon emission are individually ultraviolet divergent, with mixed
ultraviolet infrared divergent terms, the ultraviolet divergences cancel in the
sum:
\begin{eqnarray}\label{JADErealOS}
&&\frac{1}{\sigma_0}\sigma_{\JADE}^{R} \nn
&=& \frac{\alpha_s C_F}{2 \pi} \left( 2 \ln \frac{p_1^2}{Q^2}
\ln\frac{p_2^2}{Q^2} +\frac{3}{2} \ln \frac{p_1^2}{Q^2} +\frac{3}{2} \ln
\frac{p_2^2}{Q^2} \right) + \dots \ .\nn
 \end{eqnarray}
This is the same cancellation which occurs at the one-loop level in SCET
\cite{Bauer:2000ew}, in which separately ultraviolet and infrared divergent
terms cancel in the sum of the soft and collinear graphs.

The soft and collinear sectors are also individually infrared finite for the
JADE algorithm. The soft virtual vertex correction is given by
\cite{BauerSchwartz}, and contributes equally to the two-jet rate in all
definitions
\begin{eqnarray}\label{softvirtvert}
&&{1\over\sigma_0}\sigma_V^{\soft} \nn
&=& \frac{\alpha_s C_F}{2 \pi} \left ( -\frac{2}{\epsilon^2} -
\frac{2}{\epsilon}\ln\left ( -\frac{\mu^2Q^2}{p_1^2p_2^2}\right ) - \ln^2\left (
-\frac{\mu^2Q^2}{p_1^2p_2^2}\right )\right )\nn
&&+ \dots\ .
\end{eqnarray}
The soft wavefunction renormalization graphs are zero and so the cross section
in the soft sector is given by
\begin{equation}
\frac{1}{\sigma_0}\left(\sigma_{\JADE}^{\soft} + \sigma_V^{\soft}\right)=
\frac{\alpha_s C_F}{2 \pi} \left ( -\frac{2}{\epsilon^2} -
\frac{4}{\epsilon}\ln\frac{\mu}{jQ} \right )+ \dots \ .
\end{equation}
The result is purely ultraviolet divergent and agrees with the pure dimensional
regularization calculation in \eqn{JADEsoftPDR}.  The collinear contribution is
similarly free of infrared divergences.

Second, we note that the scale at which the logarithms in the NLO $n$-collinear
rate are minimized, $\mu=\sqrt{j} Q$, determines the collinear or jet scale in
SCET, $\lambda Q$, and that without the zero-bin subtraction there is no value
of $\mu$ at which the logarithms in \eqn{jadeN} are minimized. The logarithms in
the soft rate (\ref{JADEsoftPDR}) are minimized at $\mu=j Q$, the expected soft
scale in SCET, $\lambda^2 Q$. From \fig{jadeplotsfig} we see that $j Q$ is the
relevant soft scale that emerges from the multipole expansion of the JADE phase
space constraints. However, as we shall see from the SW two-jet soft rate, this
is not universally the case. The true soft scale depends on the details of the
soft theory, which is not addressed here. Furthermore the calculation of the
leading logarithmic contribution in full QCD \cite{BrownStirling1,
BrownStirling2} shows that the resummed result is not simply given by the
exponentiation of the NLO term. It has been demonstrated that the emission of
two soft gluons with large angular separation can be combined to constitute a
third jet in the JADE clustering algorithm. These types of configurations change
the leading-logarithmic two jet fraction and spoil naive exponentiation, as the
emission of subsequent soft gluons qualitatively changes the phase space
constraints.  These configurations also involve the parametrically lower
scale $j^2 Q$ \cite{BrownStirling1},
which complicates the summing of logarithms of $j$.
However, this effect does not arise until $O(\alpha_s^2)$, which is
beyond the order to which we are working.

Finally, it is instructive to look more closely at the zero-bin subtractions in
different regions of phase space. In particular, while the $n$-collinear region of
integration naturally describes the region where the $n$-collinear quark and
gluon form a jet (see \fig{jadeplotsfig}(a)), it also includes regions where the
antiquark and the gluon, as well as the quark and the antiquark, form jets. In
order for an $n$-collinear gluon to form a jet with an $\bar{n}$-collinear
antiquark, the gluon must be soft, and as a result one would expect the entire
contribution from this region of phase space to be cancelled by the zero-bin
subtraction. Similarly, the region where the $n$-collinear quark and
$\bar n$-collinear antiquark form a jet should be cancelled by the corresponding
quark and antiquark zero-bins; however, these are subleading in $j$. We show
below that this is indeed the case at $O(\alpha_s)$.\footnote{We thank S.
Freedman for this observation.}

The region where the $n$-collinear gluon and $\bar n$-collinear quark form a jet
in the JADE algorithm is defined by the region
\begin{equation}\label{nnbarjet}
\ktp>\ptm \frac{(Q-\ptm)}{Q},\qquad 0<\ptm<j Q
\end{equation}
and integrating the differential rate (\ref{nDiffDR}) over this region gives
\begin{equation}\label{nnbarjetrate}
\frac{\alpha_s C_F}{2
\pi}\left(-\frac{1}{\epsilon^2}-\frac{2}{\epsilon}\ln\frac{\mu}{jQ}+\frac{\pi^2}
{12}-2\ln^2\frac{\mu}{jQ}\right)
\end{equation}
where, as usual, we have dropped terms subleading in $j$.
The zero-bin constraints for the same jet  are 
\begin{equation}
\ktp>\ktm, \qquad 0<\ktm<j Q
\end{equation}
and integrating the differential rate (\ref{zerobinDrate}) over this region and
expanding in $j$ gives the same result as (\ref{nnbarjetrate}). Hence this
region is entirely zero-bin and is absent from the $n$-collinear rate, thereby
reducing the combinations of partons that need to be considered.  Similarly, the
region where the quark and antiquark form a jet is 
\begin{equation}
\ktp> \frac{(Q-\ptm)^2}{Q}, \qquad Q(1-j)<\ptm<Q
\end{equation}
and integrating \eqn{zerobinDrate} over this region gives a result of order $j$,
and so the rate vanishes to the order we are working. We expect that such cancellations 
will continue beyond leading order, simplifying the combinatorics of clustering multi-gluon states.

\subsection{Sterman-Weinberg and $\kt$ Jet Definitions}
\label{swkt}

It is straightforward to repeat the calculations of the previous section for the
SW and $\kt$ jet definitions.  However, each of these algorithms introduces
additional features not present in the JADE calculation: the relevant scales are
different and in both cases the zero-bin contribution is distinct from the soft
contribution.  Furthermore, in the $\kt$ definition the soft and collinear rates
are not individually infrared safe using dimensional regularization to regulate
the ultraviolet, indicating that the rate does not factorize into well-defined
soft and collinear contributions in this scheme in SCET.

\subsubsection{\SW}

Jets in the SW definition were studied in SCET in \cite{short2jet, long2jet, Trott}.
In these papers it was argued that because the kinematic cuts on the soft 
phase space were much larger than the typical soft scale, the soft phase space integral 
should be unrestricted.  In \cite{short2jet, long2jet} this is because the scaling $\beta\sim\delta$ 
is chosen, while in \cite{Trott} $\beta$ is taken to be
of order $\delta^2$, but the soft scale is taken to be $\lqcd$.
Our results differ, as we have not assumed any relative scaling between $\beta Q$, 
$\delta Q$ and $\lqcd$, and we argue that SCET power counting uniquely requires 
the restricted soft phase space in \fig{swplotsfig}(b).   (We expect, however, that if 
$\beta\sim \delta$, SCET should be matched at a lower scale onto 
a new effective theory with unrestricted soft phase space.)

Integrating the differential cross section in \eqn{softDiffDR} over the phase
space generated by the corresponding constraints, we find
\begin{eqnarray} \label{SWsoftDR}
&&\frac{1}{\sigma_0}\sigma_{\SW}^{\soft}\nn
&=& \frac{\alpha_s C_F}{2 \pi}  \left( \frac{4}{\epsilon} \ln \delta -4 \ln^2
\delta+8\ln \delta \ln \frac{\mu}{2 \beta Q}-\frac{\pi^2}{3} \right).\nn
\end{eqnarray}
By introducing quark and anti-quark off-shellnesses as we did for the JADE
algorithm, it can be shown that the total soft contribution,
$\left(\sigma_{\SW}^{\soft}+\sigma_V^{\soft}\right)/\sigma_0$, is infrared
finite, and the $1/ \epsilon$ terms are ultraviolet divergences. The logarithms
in \eqn{SWsoftDR} cannot be minimized for any choice of $\mu$ since there is a
large $\ln \delta$ in the $1/\epsilon$ term.  (See, however, \cite{Mukhi:1982bk}
in which factorization and resummation in the SW two-jet rate were studied in
perturbative QCD.)

Integrating \eqn{nDiffDR} over the phase space given by the collinear SW
constraints, we find the na\"\i ve $n$-collinear contribution to be 
\begin{eqnarray} \label{SWnDR}
\frac{1}{\sigma_0}\tilde\sigma_{\SW}^n &=& \frac{\alpha_s C_F}{2 \pi}  \left(
\frac{1}{\epsilon} \left ( \frac{3}{2}+2 \ln 2 \beta \right ) + 3 \ln
\frac{\mu}{\delta Q}\right. \nn
&&\left.+ 2 \ln 2 \beta \ln \frac{\mu^2}{2\beta\delta^2 Q^2} + \frac{13}{2}
-\frac{2 \pi^2}{3} \right).\nn
\end{eqnarray}
Note that there is no reasonable scale $\mu$ at which all the logarithms are
minimized. We now need to subtract the $\ptm \to 0$ zero-bin of the SW
$n$-collinear contribution.  Integrating over the relevant phase space gives us
\begin{eqnarray}
\frac{1}{\sigma_0}\sigma_{\SW}^{n0}&=& \frac{\alpha_s C_F}{2 \pi}  \left (
-\frac{1}{\epsilon^2} - \frac{2}{\epsilon}\ln \frac{\mu}{2\beta\delta
Q}\right.\nn
&& \left.- 2\ln^2 \frac{\mu}{2\beta\delta Q} + \frac{\pi^2}{12}\right ).
\end{eqnarray}
The zero-bin gives a nontrivial contribution that is not equal to the soft
contribution, because the region of integration generated by taking the
collinear and then soft limit is not the same as taking the soft limit of the
QCD SW phase space. It is interesting to note that the scale in the
$n$-collinear zero-bin, $\beta \delta Q$, corresponds to the $p_\perp $ of a
parton at the edge of the cone with the maximum energy allowed outside the cone,
$\beta Q$. This corresponds to the intersection point of \fig{swplotsfig}(c),
generated by a consistent expansion of phase space constraints in the effective
theory.

The zero-bin subtracted result for the $n$-collinear sector is
\begin{eqnarray}\label{sw_ncoll_zerobinsub}
\frac{1}{\sigma_0}(\tilde\sigma_{\SW}^{n}-\sigma_{\SW}^{n0})&=& \frac{\alpha_s
C_F}{2 \pi} \left(  \frac{1}{\epsilon^2}+\frac{3}{2 \epsilon}
+\frac{2}{\epsilon} \ln \frac{\mu}{\delta Q}\right.\nn
&+& \left.  3 \ln \frac{\mu}{\delta Q} +2 \ln^2 \frac{\mu}{\delta Q} - \frac{3
\pi^2}{4}+\frac{13}{2} \right)\nn
\end{eqnarray}
where the logarithms are now minimized at $\mu = \delta Q$, unlike in
\eqn{SWnDR}. The collinear scale, $\delta Q$, corresponds to the $p_\perp $ of a
parton at the edge of the cone with typical collinear energy $O(Q)$. The
emission of a collinear gluon in the $\bar{n}$ direction, i.e. from the
anti-quark, gives the same result. 

The $n$-collinear rate is independent of the jet parameter $\beta$, because the
phase space region in \fig{swplotsfig}(b) with a collinear gluon outside the
cone with energy less than $\beta Q$, where $\beta \ll 1$, corresponds to the
zero-bin. This contribution is entirely removed by the zero-bin subtraction and
\eqn{sw_ncoll_zerobinsub} is given only by the region where the $n$-collinear
quark and gluon lie in the cone. This underscores the consistency of the phase
space expansion in Section \ref{ps} and the zero-bin prescription. The soft
sector resolves the cone in addition to the scale $\beta Q$ and gives rise to
the double logarithm cross term in the SW result below.

Combining these results gives 
\begin{eqnarray}\label{sw_2jet_rate}
f^{\SW}_2&=& \frac{|C_2|^2}{|Z_2|^2}\left(1 +
{2\over\sigma_0}(\tilde\sigma_{\SW}^{n}-\sigma_{\SW}^{n0})+{1\over\sigma_0}
\sigma_{\SW}^{\soft}\right) \nn
&=& 1 + \frac{\alpha_s C_F}{\pi} \left( -4 \ln 2 \beta \ln \delta-3\ln \delta
-\frac{\pi^2}{3}+\frac{5}{2} \right)\nn
\end{eqnarray}
in agreement with the full QCD calculation \cite{SW}.

\subsubsection{$\kt$}

The $\kt$ two-jet rate in SCET reveals a more subtle cancellation of divergences
than the previous two algorithms and highlights again the importance of zero-bin
subtractions.  Integrating the differential cross section for the emission of a
soft gluon over the soft phase space in \fig{ktplotsfig}(b), we find that
$\sigma^{\soft}_{\kt}$ is not regulated in dimensional regularization. 
Performing the $\ktp$ integral first over the $\bar{q} g$ jet region of phase
space, we obtain a term proportional to
\begin{equation}\label{ktdiverge}
\frac{ d \sigma^{\soft}_{\kt}}{d \ktm} \propto \frac{\left(Q^2
y_c-({\ktm})^2\right)^{-\epsilon}}{\epsilon \ \ktm}+\cdots,
\end{equation}
where the ellipses denote terms which are finite in $d=4-2 \epsilon$ dimensions.
This term causes the $\ktm$ integration to diverge at zero. A similar problem
arises when integrating over the soft region generated by the $q g$ jet
constraint. Despite this divergence, the total two-jet cross section is finite
in QCD and so must be finite in SCET. The region that gives rise to this
divergence is also integrated over in the zero-bin calculations and since the
soft and zero-bin integrands are the same the divergence cancels in the
difference. Integrating the soft differential rate over the combined soft and
zero-bin regions gives a finite result in $d$ dimensions:
\begin{eqnarray}
&&\frac{1}{\sigma_0}(\sigma_{\kt}^{\soft}-\sigma_{\kt}^{n0}-\sigma_{\kt}^{\bar{n
}0}) \nn
&=& \frac{\alpha_s C_F}{2\pi} \left( \frac{2}{\epsilon^2} +\frac{2}{\epsilon}
\ln \frac{\mu^2}{y_c Q^2}+\ln^2\frac{\mu^2}{y_c Q^2}-\frac{\pi^2}{3} \right)\nn
\end{eqnarray}
where we see the soft scale $\sqrt{y_c} Q$ appear as in \fig{ktplotsfig}. We
combine this with the rate to produce an $n$-collinear gluon,
\begin{eqnarray}
\frac{1}{\sigma_0}\tilde\sigma_{\kt}^{n}&=&\frac{\alpha_s C_F}{2\pi} \left(
\frac{1}{\epsilon} \left(\frac{3}{2}+\ln y_c \right)+\ln \frac{\mu^2}{y_c Q^2}
\left(\frac{3}{2}+\ln y_c \right)\right.\nn
&&-\left. 3\ln2-\frac{\pi^2}{3}+\frac{7}{2} \right) 
\end{eqnarray} 
to obtain the total two-jet rate for emission of a real gluon
\begin{eqnarray}\label{ktreal}
&&\frac{1}{\sigma_0}(\tilde\sigma_{\kt}^n+\tilde\sigma_{\kt}^{\bar{n}}+
\sigma_{\kt}^\soft-\sigma_{\kt}^{n0}-\sigma_{\kt}^{\bar{n}0}) \nn
&=& \frac{\alpha_s C_F}{2\pi} \left( \frac{2}{\epsilon^2}
+\frac{1}{\epsilon}\left(2 \ln \frac{\mu^2}{Q^2}+3\right)+\ln^2\frac{\mu^2}{Q^2}
 \right.  \nn 
&& +\left. 3 \ln\frac{\mu^2}{Q^2}-\ln^2 y_c-3 \ln y_c-6 \ln2 -\pi^2+7 \right)\nn
\end{eqnarray}
where again $n$ and $\bar{n}$ collinear gluon emission give the same
contribution and the virtual piece vanishes. Including the counter-term $Z_2$
and the Wilson coefficient $C_2$, we reproduce the known NLO $\kt$ result
\cite{BrownStirling2}
\begin{eqnarray}
&&f_2^{\kt}\nn
&=& 1 + \frac{\alpha_s C_F}{2\pi} \left(-\ln^2 y_c-3 \ln
y_c-6\ln2+\frac{\pi^2}{6}-1\right).\nn
\end{eqnarray}
This calculation re-emphasizes the importance of zero-bin subtraction: without
it, the evaluation of a finite $f_2^{\kt}$ would not be possible. Since the soft
and collinear cross sections are not regulated in dimensional regularization, it
is useful to regulate the infrared and ultraviolet divergences separately by
taking the outgoing quark and antiquark off-shell. The resulting rate for soft
gluon emission then becomes
\bea\label{kTsoftOS}
\frac{1}{\sigma_0}\sigma_{\kt}^{\soft} = \frac{\alpha_s C_F}{2 \pi} \ln^2 \frac{
p_1^2 p_2^2 }{Q^4\, y_c} + \dots \ .
\eea
Note that unlike the previous algorithms, the soft real emission result is not
ultraviolet divergent. Combining this with the contribution from the soft
virtual vertex correction (\ref{softvirtvert}) gives
\begin{eqnarray}
&&\frac{1}{\sigma_0}\left(\sigma_{\kt}^{\soft} + \sigma^\soft_V\right) \nn
&= &\frac{\alpha_s C_F}{2 \pi} \left( -\frac{2}{\epsilon^2}
-\frac{2}{\epsilon}\ln\frac{\mu^2Q^2}{p_1^2p_2^2}
+2\ln\frac{p_1^2p_2^2}{Q^4}\ln\frac{\mu^2}{y_cQ^2} \right )\nn
&&+\dots .
\end{eqnarray}
This shows explicitly that the rate in the soft sector is not infrared safe.

The rate for $n$-collinear gluon emission and the zero-bin are, respectively,
\begin{eqnarray}
&&\frac{1}{\sigma_0}\tilde\sigma_{\kt}^{n} \nn
& = & \frac{\alpha_s C_F}{2\pi} \left(
-\frac{2}{\epsilon^2}-\frac{2}{\epsilon}\ln \frac{\mu^2}{p_1^2
\sqrt{y_c}}-\ln^2\frac{\mu^2}{p_1^2}+\frac{3}{2} \ln \frac{p_1^2}{Q^2 y_c}
\right) \nn&&+\dots  \nn
&&\frac{1}{\sigma_0}\sigma_{\kt}^{n0} \nn
&=& \frac{\alpha_s C_F}{2\pi} \left( -\frac{2}{\epsilon^2}-\frac{2}{\epsilon}\ln
\frac{\mu^2}{p_1^2
\sqrt{y_c}}+\ln^2\frac{p_1^2}{y_cQ^2}-\ln^2\frac{\mu^2}{p_1^2} \right) \nn && +
\dots \  .
\eea
and their difference gives us the zero-bin subtracted result
\begin{eqnarray}\label{kTnczbsOS}
\frac{1}{\sigma_0}\sigma_{\kt}^{n}=\frac{\alpha_s C_F}{2\pi} \left(
-\ln^2\frac{p_1^2}{y_c \, Q^2}+\frac{3}{2}\ln \frac{p_1^2}{y_c \, Q^2} \right) +
\dots \ .\nn
\eea
As with the soft sector, the phase-space integration for the $n$-collinear real
emission is ultraviolet finite but infrared divergent.
Combining the real emission contributions to the $\kt$ two-jet cross section, we
find
\begin{eqnarray}\label{kTrealOS}
\frac{1}{\sigma_0}\sigma_{\kt}^R&=&\frac{1}{\sigma_0}(\sigma_{\kt}^{n} +
\sigma_{\kt}^{\bar{n}} +\sigma_{\kt}^{\soft}) \nn 
&=& \frac{\alpha_s C_F}{2\pi} \left(
\frac{3}{2}\left(\ln\frac{p_1^2}{Q^2}+\ln\frac{p_2^2}{Q^2}\right)+2\ln\frac{p_1^
2}{Q^2}\ln\frac{p_2^2}{Q^2} \right)\nn 
&&+ \dots \ .
\end{eqnarray}
The infrared divergences in \eqn{kTrealOS} are completely cancelled by the total
virtual contribution $\sigma_V$ given in \eqn{virt}. As expected, the virtual
graphs convert the infrared divergences in the real emission diagrams into
ultraviolet ones. While SCET reproduces the known NLO $\kt$ result, the soft and
collinear rates are not independently infrared safe, indicating for the $\kt$
phase space the soft and collinear modes do not factorize in SCET using
dimensional regularization to regulate the ultraviolet.

\section{Factorization and Scheme-Dependence}

It is useful to examine the failure of SCET to factorize the $\kt$ rate into
separately infrared safe soft and collinear pieces, particularly given the fact
that the regions of integration for the soft gluons are quite similar in the
infrared between $\kt$ and $\JADE$.  Instead, the bad behaviour in
\eqn{ktdiverge} comes from the region of large $k^+$ and small $k^-$ and
vice-versa - a region which is infrared divergent, but sensitive to the
ultraviolet regulator.   Since, as we have shown, the ultraviolet divergences in
the phase space integrals cancel between the soft and collinear degrees of
freedom, this is an unphysical region, and so cancels from the total rate.  The
same cancellation occurs at the one-loop level, in which terms of order
$1/\epsilon_{\rm UV} \ln p_i^2$ cancel between soft and collinear graphs
\cite{Bauer:2000ew}.   However, this unphysical region can also be eliminated by
defining the soft function with a cutoff $\lambdaf$.    In particular, we show
in this section that while the $\kt$ algorithm in dimensional regularization
does not factorize in SCET into separate infrared safe contributions, regulating
the ultraviolet with a cutoff on the light-cone components of the gluon
momentum, 
\begin{equation}\label{cutoff}
|k^+|<\Lambda_f,\ \ |k^-|<\Lambda_f
\end{equation}
results in an infrared safe soft function.

Integrating the soft rate over the relevant region for $\kt$, including the
cutoff (\ref{cutoff}), and continuing to work in $d$ dimensions to regulate the
infrared, we find for real soft gluon emission
\begin{eqnarray}\label{kTprimesoft}
\frac{1}{\sigma_0}\sigma_{\kt}^{\soft}&=&\frac{\alpha_s C_F}{2 \pi} \left(
\frac{2}{\epsilon^2} +\frac{2}{\epsilon} \ln
\frac{\mu^2}{\lambdaf^2}-\ln^2\frac{y_c Q^2}{\lambdaf^2}\right.\nn
&&+\left.\ln^2\frac{\mu^2}{\lambdaf^2}-\frac{\pi^2}{3}    \right)\nn
\end{eqnarray}

Similarly, the same regulator for soft real gluon emission in the JADE algorithm
gives
\begin{eqnarray}\label{JADEprimesoft}
\frac{1}{\sigma_0}\sigma_{\JADE}^{\soft}
&=& \frac{\alpha_s C_F}{2 \pi} \left( \frac{2}{\epsilon^2} +\frac{2}{\epsilon}
\ln \frac{\mu^2}{\lambdaf^2}-\frac12\ln^2\frac{j ^2 Q^2}{\lambdaf^2}\right.\nn
&&+\left.\ln^2\frac{\mu^2}{\lambdaf^2}-\frac{\pi^2}{6}    \right)\nn
\end{eqnarray}

Note that with a cutoff, the $1/\epsilon^2$ and Sudakov double logs $\ln^2j$ and
$\ln^2 y_c$ are entirely contained within the soft function, as opposed to pure
dimensional regularization, in which the collinear graphs also contain double
logs.  This is in agreement with \cite{BrownStirling1,BrownStirling2}, where the
Sudakov logs are calculated entirely from the soft graphs.

The soft virtual vertex correction with a cutoff of $\Lambda_f$ in $|k^+|$ and
$|k^-|$ gives a modified vertex correction
\begin{equation}
\sigma_V^{\soft} = \frac{\alpha_s C_F}{2 \pi} \left ( -\frac{2}{\epsilon^2} -
\frac{2}{\epsilon}\ln\frac{\mu^2}{\Lambda_f^2} - \ln^2 \frac{\mu^2}{\Lambda_f^2}
+ \frac{\pi^2}{6}\right )
\end{equation}
giving the finite results
\begin{eqnarray}
\frac{1}{\sigma_0}\left(\sigma_{\kt}^{\soft} +
\sigma_V^{\soft}\right)&=&-\frac{\alpha_s C_F}{2 \pi}\left(\ln^2{y_c Q^2\over
\Lambda_f^2}+\frac{\pi^2}{6}\right).\nn
\frac{1}{\sigma_0} \left( \sigma_{\JADE}^{\soft} + \sigma_V^{\soft} \right)&=&
-\frac{\alpha_s C_F}{4 \pi}\ln^2{j^2 Q^2\over \Lambda_f^2}.
\end{eqnarray}

Note that the infrared divergences cancel between the real and virtual graphs,
and that there are no large logs in the soft function for $\Lambda_f$ of order
the relevant soft scale, $jQ$ or $\sqrt{y_c} Q$. 

These results demonstrate the fact that factorization of rates in SCET into soft and collinear
components is scheme-dependent. Such dependence on infrared regulators was also noted 
in a different context in \cite{Hornig:2009vb} and \cite{Chiu:2009mg}. Using the method 
introduced in \cite{Hornig:2009vb} to test infrared safety at one loop, one would conclude that 
the soft contribution to the $\kt$ rate is infrared divergent. This differs from our results, because, 
as we have shown, the infrared safety of the soft function is ultraviolet regulator dependent. 
Introducing a cutoff removes the unphysical region of $k^\pm \to 0$ and 
$k^\mp \to \infty$ and results in an infrared safe soft contribution to the 
two-jet $\kt$ rate.\footnote{Similarly, the NLO soft function for angularities, $\tau_a$, for $1<a<2$ integrated 
over $\tau_a$ between $0$ and $1$ can be shown to be infrared finite if defined 
with an ultraviolet cutoff.} The bad behaviour of $\kt$ in dimensional regularization in 
SCET is therefore a feature of dimensional regularization, not of SCET. The factorization for 
jet rates depends on the ultraviolet regulator of the theory as well as the infrared.

\section{Conclusion}

We have presented a consistent treatment of phase-space integrals over soft and
collinear degrees of freedom in SCET, illustrating this with the explicit
example of the NLO dijet rate for three different jet algorithms. In this
approach the phase space for different modes in the effective theory are
insensitive to details above their cutoff, giving real emission contributions
with ultraviolet divergences which cancel between the collinear and soft
sectors. Although the leading order SCET Lagrangian separates soft and collinear
modes and the differential cross section has been shown to factorize, we
demonstrated that using dimensional regularization the $\kt$ algorithm does
not factorize into infrared safe soft and collinear rates.  We showed that this is
related to a divergence in an unphysical region which cancels between the soft
and collinear sectors, and is sensitive to the ultraviolet regulator.

Zero-bin subtraction is necessary to consistently integrate over the phase space
configurations that need to be considered in a given jet algorithm. The zero-bin
subtraction was shown to entirely remove regions of the na\"\i ve collinear rate
where $n$ and $\bar{n}$ collinear degrees of freedom form a jet at NLO in the
JADE algorithm and for collinear partons outside the cone in SW. The $\kt$ and
SW dijet rates provide nontrivial examples of zero-bin subtraction, which are
different from the soft contribution. 

We have not attempted to sum logarithms of the small jet parameters at this
stage.  While the running of $C_2$ makes summing some of the logarithms
straightforward, the soft physics in these theories is more complicated.  For
example, the JADE algorithm is known not to exponentiate:  there are three-jet
configurations which contribute at $O(\alpha_s^2\ln^4 j)$ in which two gluons,
which would na\i\" vely be unresolved from the quarks, are combined to form a
third jet \cite{BrownStirling1}.  Such configurations have no simple relation to
the one-gluon phase space and are not obtained by exponentiating the one-loop
result.  From an effective field theory viewpoint, these configurations also
involve the scale $j^2 Q$, which is parametrically smaller than the soft scale
$jQ$.  The soft function for the SW algorithm, in contrast, na\"\i vely has an
anomalous dimension of order $\ln\delta$, and so large logarithms of $\delta$
cannot be resummed in this formulation of the low-energy theory.

\acknowledgements

We thank A. Blechman, C. Bauer, S. Freedman, Z. Ligeti, A. Manohar, I. Rothstein and 
M. Trott for useful discussions and comments on the manuscript. This work was 
supported by the Natural Sciences and Engineering Research Council of Canada.

\begin{appendix}

\section{Offshell calculations}\label{appendixa}

The SCET differential cross section for soft gluon emission and offshell quarks,
$p_1^2, p_2^2\neq 0$, is 
\begin{eqnarray}
\frac{1}{\sigma_0}d \sigma^{\soft} &=& \frac{\alpha_s C_F}{2 \pi}  \frac{\mu^{2
\epsilon} e^{\epsilon \gamma_E} }{\Gamma(1-\epsilon)} \, \theta (\ktp \ktm) \,
d\ktp d\ktm \nn
&&\times \frac{2 \, Q^2 \, (\ktp \ktm)^{- \epsilon}}{(Q \ktp + p_1^2)(Q \ktm +
p_2^2)},
\end{eqnarray}
where $p_1^2=Q k_1^+$,  $p_2^2=Q k_2^-$ and $p_3^2=0$.
The JADE two-jet constraints become
\begin{eqnarray}\label{JADEsoftOS}
\frac{M_{13}^2}{Q^2} &=& \frac{Q \ktp + p_1^2}{ Q^2} < j , \quad
\frac{M_{23}^2}{Q^2} = \frac{Q \ktm + p_2^2}{ Q^2} < j ,\nn
\frac{M_{12}^2}{Q^2} &=& 1
\end{eqnarray}
and integrating over the soft phase space gives
\begin{eqnarray}\label{JADEsigmaSOS}
\frac{1}{\sigma_0}\sigma_{\JADE}^{\soft} &=& \frac{\alpha_s C_F}{2 \pi} \left(
\frac{1}{\epsilon} \left( 4 \ln j -2 \ln \frac {p_1^2}{Q^2}-2 \ln
\frac{p_2^2}{Q^2} \right) \right.\nn
&&+\left(\ln \frac {p_1^2}{Q^2} + \ln \frac{p_2^2}{Q^2} \right)^2\nn
&&-\left. 2 \left(\ln \frac {p_1^2}{Q^2} + \ln \frac{p_2^2}{Q^2} \right) \ln
\frac{\mu^2}{Q^2}\right)+\cdots
\end{eqnarray}
where the ellipses denote finite constant terms.

Similarly, the SCET differential cross section for $n$-collinear gluon emission
with off-shellness is  
\begin{eqnarray}\label{ndiffOS}
\frac{1}{\sigma_0}d \sigma^{n} &=& \frac{\alpha_s C_F}{2 \pi}  \frac{\mu^{2
\epsilon} e^{\epsilon \gamma_E} }{\Gamma(1-\epsilon)} d\ktp d\ptm (\ptm \ktp)^{-
\epsilon} \nn
&&\times \left( \frac{(1-\epsilon)\, \ptm \ktp}{(p_1^2+Q \ktp)^2} + \frac{2
(Q-\ptm)}{\ptm (p_1^2+Q \ktp)} \right)\nn
\end{eqnarray}
and the corresponding JADE two-jet constraints  are 
\begin{eqnarray}
\frac{M_{13}^2}{Q^2}& =&\frac{Q \ktp +p_1^2}{Q (Q-\ptm)} < j ,
\quad\frac{M_{23}^2}{Q^2} = \frac{Q \ptm +p_2^2}{ Q^2} < j, \nn 
\frac{M_{12}^2}{Q^2} &=& \frac{Q (Q-\ptm)+p_1^2+p_2^2}{Q^2}<j.
\end{eqnarray}
Note that the off-shellnesses in $M_{23}^2$ and $M_{12}^2$ are suppressed with
respect to the label momenta and thus can be dropped. Integrating \eqn{ndiffOS}
over the phase space given by these constraints, we find
\begin{eqnarray}\label{jadeNos}
&&\frac{1}{\sigma_0}\tilde\sigma_{\JADE}^{n } \nn
&=& \frac{\alpha_s C_F}{2 \pi} \left(-\frac{2}{\epsilon^2}+  \frac{1}{\epsilon}
\left( 2 \ln j+2 \ln \frac{p_1^2}{Q^2}-2 \ln \frac{\mu^2}{Q^2} \right) \right.
\nn
& &-\left. \ln^2 \frac{p_1^2}{Q^2} + 2 \ln\frac{\mu^2}{Q^2}\ln \frac{p_1^2}{Q^2}
+ \frac{3}{2} \ln \frac{p_1^2}{Q^2}\right) + \dots\ .\nn
\end{eqnarray}

The $\ptm \to 0$ zero-bin for the $n$-collinear differential cross section is
obtained from \eqn{ndiffOS} by taking the soft limit: 
 \bea\label{ndiffZeroOS}
\frac{1}{\sigma_0}d \sigma^{n0}&=& \frac{\alpha_s C_F}{2 \pi}  \frac{\mu^{2
\epsilon} e^{\epsilon \gamma_E} }{\Gamma(1-\epsilon)} d\ktp d\ptm (\ptm \ktp)^{-
\epsilon}\nn
&&\times \frac{2 \, Q}{\ptm (p_1^2+Q \ktp)}.
\end{eqnarray}
The JADE constraints for this zero-bin are the same as the soft ones in
\eqn{JADEsoftOS}. Performing the phase space integration gives
\bea \label{jadeN0os}
\frac{1}{\sigma_0}\sigma_{\JADE}^{n0} = \frac{\alpha_s C_F}{2 \pi}
\left(-\frac{2}{\epsilon^2}-  \frac{2}{\epsilon} \ln \frac {\mu^2}{j^2 Q^2}
\right) +\dots \ .
\eea

The zero-bin subtracted result, which is the difference betweem \eqn{jadeNos}
and \eqn{jadeN0os}, is not particularly illuminating. It should be noted,
however, that this zero-bin subtraction gets rid of the $1/\epsilon^2$ term,
which is also absent in the contribution from soft gluon emission in
\eqn{JADEsigmaSOS}. Thus the total contribution from real gluon emission is free
of such terms. The result for  $\bar{n}$-collinear gluon emission will be the
same as that for $n$-collinear gluon emission with $p_1^2 \to p_2^2$.
Combining the real emission contributions to the JADE cross section gives
\begin{eqnarray}\label{app:JADErealOS}
&&\frac{1}{\sigma_0}\sigma_{\JADE}^{R} \nn
&=&\frac{1}{\sigma_0}\left( (\tilde\sigma_{\JADE}^{n}-\sigma_{\JADE}^{n0})+
(\tilde\sigma_{\JADE}^{\bar{n}}-\sigma_{\JADE}^{\bar{n}0})+\sigma_{\JADE}^{\soft
} \right) \nn
 &=& \frac{\alpha_s C_F}{2 \pi} \left( 2 \ln \frac{p_1^2}{Q^2}
\ln\frac{p_2^2}{Q^2} +\frac{3}{2} \ln \frac{p_1^2}{Q^2} +\frac{3}{2} \ln
\frac{p_2^2}{Q^2} \right) + \dots \ . \nn
 \eea
Notice that this result is free of ultraviolet divergences, and off-shellness is
regulating all of its infrared divergences. The collinear and the soft sectors are
individually ultraviolet divergent, but these ultraviolet divergences arising from the
phase space cancel completely with one another in the sum.

With off-shellness, the virtual diagrams are no longer zero, and they have been
previously calculated with off-shellness, for example in \cite{dis} for deep inelastic
scattering and
in \cite{BauerSchwartz} for $e^+e^-$ annihilation. The zero-bin subtractions of
the collinear virtual graphs also vanish with this regulator \cite{zerobin,
DeltaRegulator}. At the amplitude level, we sum up all the virtual vertex
corrections and subtract half the wavefunction renormalization for each external
(anti-)quark:
\bea \label{virt}
I_{V} &=& \frac{\alpha_s C_F}{4 \pi} \left(\frac{2}{\epsilon^2} +
\frac{3}{\epsilon} - \frac{2}{\epsilon}\ln\frac{-Q^2}{\mu^2} - 2 \ln
\frac{p_1^2}{Q^2} \ln \frac{p_2^2}{Q^2}\right.\nn
&&-\left.\frac{3}{2}\ln \frac{p_1^2}{Q^2}-\frac{3}{2}\ln \frac{p_2^2}{Q^2}
\right)+ \dots\ .
\end{eqnarray}
The virtual graphs' contribution to the two-jet rate is $\sigma_V=2Re(I_V)$. We
can then see that the IR divergences from real gluon emission in
\eqn{app:JADErealOS} will be completely cancelled by the virtual contributions,
and the UV divergent terms in $\sigma_V$ will be cancelled by the counter term
$|Z_2|^2$.

We can also focus on the soft sector to investigate its IR safety. The soft
virtual vertex correction is given by \cite{BauerSchwartz}:
\begin{eqnarray}
&&I_V^{\soft} \nn&=& \frac{\alpha_s C_F}{4 \pi} \left ( -\frac{2}{\epsilon^2} -
\frac{2}{\epsilon}\ln\left ( -\frac{\mu^2Q^2}{p_1^2p_2^2}\right ) - \ln^2\left (
-\frac{\mu^2Q^2}{p_1^2p_2^2}\right )\right ) \nn
&&+\dots\ .
\end{eqnarray}
The soft wavefunction renormalisation graphs are zero, so in the soft sector,
the soft virtual vertex correction and the soft gluon bremstrahhlung are the
only two diagrams we need to add:
\begin{equation}
\frac{1}{\sigma_0}\left(\sigma_{\JADE}^{\soft} + \sigma_V^\soft \right)=
\frac{\alpha_s C_F}{2 \pi} \left ( -\frac{2}{\epsilon^2} -
\frac{4}{\epsilon}\ln\frac{\mu}{jQ} \right )+ \dots\ .
\end{equation}
This agrees with our pure dimensional regularization calculation in
\eqn{JADEsoftPDR}. This also shows that the rate in the soft sector is infrared
finite. The collinear contribution is also IR safe because the sum of all
sectors is free of infrared divergences.

$\kt$:  The $\kt$ phase space regions shown in Table \ref{jettable} are not
affected by the introduction of the off-shellnesses, with the only exception
that the constraint
\begin{equation}
\vphantom{\Bigg(} \mathrm{min} \left(\frac{\ktp}{\ptm}, \frac{\ktp
\ptm}{(Q-\ptm)^2}\right)<y_c 
\end{equation}
is slightly modified to
\begin{equation}
\vphantom{\Bigg(} \mathrm{min}\left(\frac{Q-\ptm}{\ptm},
\frac{\ptm}{Q-\ptm}\right) \frac{Q^2 \ktp +\ptm p_1^2}{Q^2 (Q-\ptm)} < y_c.
\end{equation}
The calculation is otherwise straightforward.

\end{appendix}

\end{document}